\documentclass{iaus}
\usepackage{graphicx}

\newcommand{\bc}{\begin{center}}
\newcommand{\ec}{\end{center}}
\newcommand{\ben}{\begin{enumerate}}
\newcommand{\een}{\end{enumerate}}
\newcommand{\bei}{\begin{itemize}}
\newcommand{\eei}{\end{itemize}}
\newcommand{\be}{\begin{equation}}
\newcommand{\ee}{\end{equation}}
\newcommand{\ber}{\begin{eqnarray}}
\newcommand{\eer}{\end{eqnarray}}

\title[IAUS273~~Magnetohydrostatic Equilibrium in Starspots] 
{Magnetohydrostatic equilibrium in starspots: dependences on color ($T_{eff}$) and surface gravity ($g$)}

\author[Rajaguru and Hasan]   
{S.P. Rajaguru and S.S. Hasan}

\affiliation{Indian Institute of Astrophysics, Bangalore - 560034, India, \\ 
email: {\tt rajaguru@iap.res.in}} 

\pubyear{2010}
\volume{273}  
\pagerange{119--126}
\jname{Physics of Sun and Star Spots}
\editors{D.P. Choudhary, \& K. Strassmeier, eds.}
\begin{document}

\maketitle

\begin{abstract}
Temperature contrasts and magnetic field strengths of sunspot umbrae broadly follow the thermal-magnetic
relationship obtained from magnetohydrostatic equilibrium. Using a compilation of recent
observations, especially in molecular bands, of temperature contrasts of starspots in cool stars, and
a grid of Kurucz stellar model atmospheres constructed to cover layers of sub-surface convection
zone, we examine how the above relationship scales with effective temperature ($T_{eff}$), surface
gravity $g$ and the associated changes in opacity of stellar photospheric gas. We calculate expected
field strengths in starpots and find that a given relative reduction in temperatures (or the same darkness
contrasts) yield increasing field strengths against decreasing $T_{eff}$ due to a combination of pressure
and opacity variations against $T_{eff}$.
\keywords{Sun: magnetic fields, sunspots, stars: magnetic fields, stars: spots, stars: activity}
\end{abstract}
\section{Introduction}

Despite a lack of deductive magnetohydrodynamic explanation for the
formation and the equilibrium of a sunspot, extensive observations in
combination with magnetohydrostatic models have provided reasonable
understanding of the thermal-magnetic structure of sunspots in the observable layers.
Reduced temperature and gas pressure inside sunspots dictate dominantly the thermal-magnetic 
relationship derived from the magneto-hydrostatic balance. Such a relationship
causes the Wilson depression -- geometrical
depression of the observable optical depth unity level within a sunspot --, which provides
an explanation for the field strengths observed in sunspots and also relates the 
intensity (or the brightness) contrasts to field strengths.
Here we examine how such a thermal-magnetic relationship scales with the stellar
parameters, viz. the effective temperature $T_{eff}$ and surface gravity $g$ as well
as the associated changes in the opacity of the stellar photospheric gas.
We then discuss the implications of such scalings for the interpretations of observed
field strengths. We also discuss how such scalings could be crucial players in the activity related
photospheric brightness variations and their correlation with other activity
measures. 

\section{Method of Calculation}
Thermal-magnetic relationship for starspots is obtained from a 
simplified magnetohydrostatic (MHS) condition, that relates the magnetic field strength
to the temperature at the axis of a vertical column of magnetic field (\cite[Solanki \etal\ (1993)]{solankietal93}).
The radial component of the magneto-hydrostatic force-balance equation (Maltby 1977, Solanki
et al. 1993), after neglecting the magnetic curvature force and with the use of the equation of state
$P=R\rho T/\mu$, yields the thermal-magnetic relation,
\be
\frac{T(r,z)}{T_{e}(z)}=\frac{\mu(r,z)\rho_{e}(z)}{\mu_{e}(z)\rho(r,z)}\left [1-\frac{B_{z}^{2}(r,z)}{8\pi P_{e}(z)}
\right ],
\label{eqn1}
\ee
where $P, T, \rho, \mu$ and $R$ denote the gas pressure, temperature, density, mean molecular weight and
the gas constant respectively. The subscript $e$ stands for the external atmosphere. $B_{z}$ is the
vertical component of magnetic field and $r$ is the radial distance from the center of the spot and
$z$ is the depth measured from continuum optical depth $\tau_{c,e}$=1. We further neglect the $r$-dependence of quantities
in the above equation, thus the calculated quantities refer to the axis of the spot, and we refer $B_{z}$ simply as $B$ hereafter.
Eqn.\ref{eqn1} is valid for each level $z$. The variation of external atmospheric quantities
with depth $z$ are determined by $g$ and $T_{eff}$ of the parent star and are taken from the
grid of Kurucz stellar models constructed to cover deeper regions of the convection zone using the ATLAS9
stellar atmosphere code (Kurucz 2001). 
We prescribe temperature contrasts for starspots under two cases, case (i): use the empirical
relation $\Delta T = (590.*log g) - 680 K$ (\cite{onealetal96}) to determine the effective temperature of the spot
$T_{eff,spt}$ in a parent star characterised by $g$ and $T_{eff}$:
$T_{spt}(\tau_{c}=2/3)=T_{eff,spt}=T_{eff}-\Delta T $, case (ii):
the temperature contrasts are independent of $g$ and are of
a constant ratio of $T_{eff}$: $\Delta T = 0.3 T_{eff}$ (\cite{berdyugina05}).
\begin{figure}[b]
\begin{center}
 \includegraphics[width=3.5in]{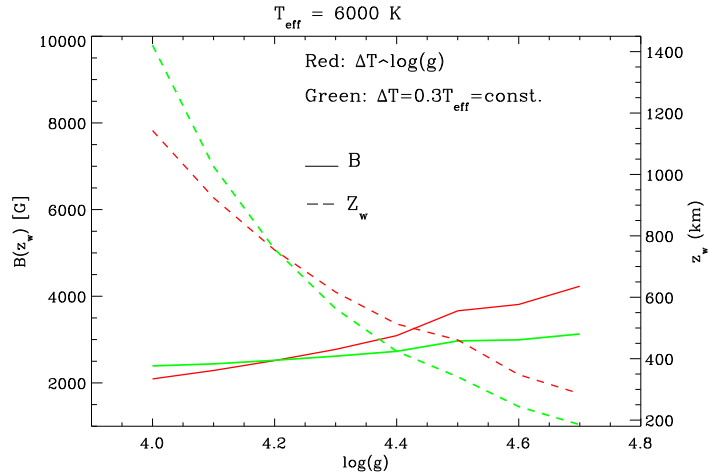}
 \caption{Variation of field strength $B$ (solid lines)  and the Wilson depression $Z_{w}$ (dashed lines) against 
$log (g)$ for $T_{eff}=6000 K$}
   \label{fig1}
\end{center}
\end{figure}
Hydrostatic equilibrium inside the spot yields, to a very good approximation (Cox and Giuli 1968),
\be
P(\tau_{c}=2/3)=\frac{2}{3}\frac{g}{\kappa_{R} (\rho,T_{eff,spt})} .
\label{eqn2}
\ee
The density inside the spot at $\tau_{c}=2/3$ is determined by solving the above equation
with the use of Rosseland mean opacities $\kappa_{R}$ from the tables of Kurucz (1993) and Alexander
and Ferguson (1994). Saha's equation is used to determine the mean molecular weight.
The density difference between the $\tau_{c}=2/3$ and $\tau_{c}=1$ levels within the spot is
assumed to be negligible, i.e., $\rho (\tau_{c}=1$) $\approx$ $\rho (\tau_{c}=2/3)$.
Using the assumption that $\rho (\tau_{c}=1)=\rho_{e}(z=Z_{w})$, i.e., the densities
inside and outside the spot are equal at the level of Wilson depression $Z_{w}$ (\cite{maltby77}), we find the depth $z$
at which the above equality is satisfied in the parent stellar model. This gives the value of $Z_{w}$, and
now we determine the only unknown quantity $B$ at this level using Eqn.\ref{eqn1}.

\section{Results and Discussions}
\begin{figure}[b]
\begin{center}
 \includegraphics[width=3.5in]{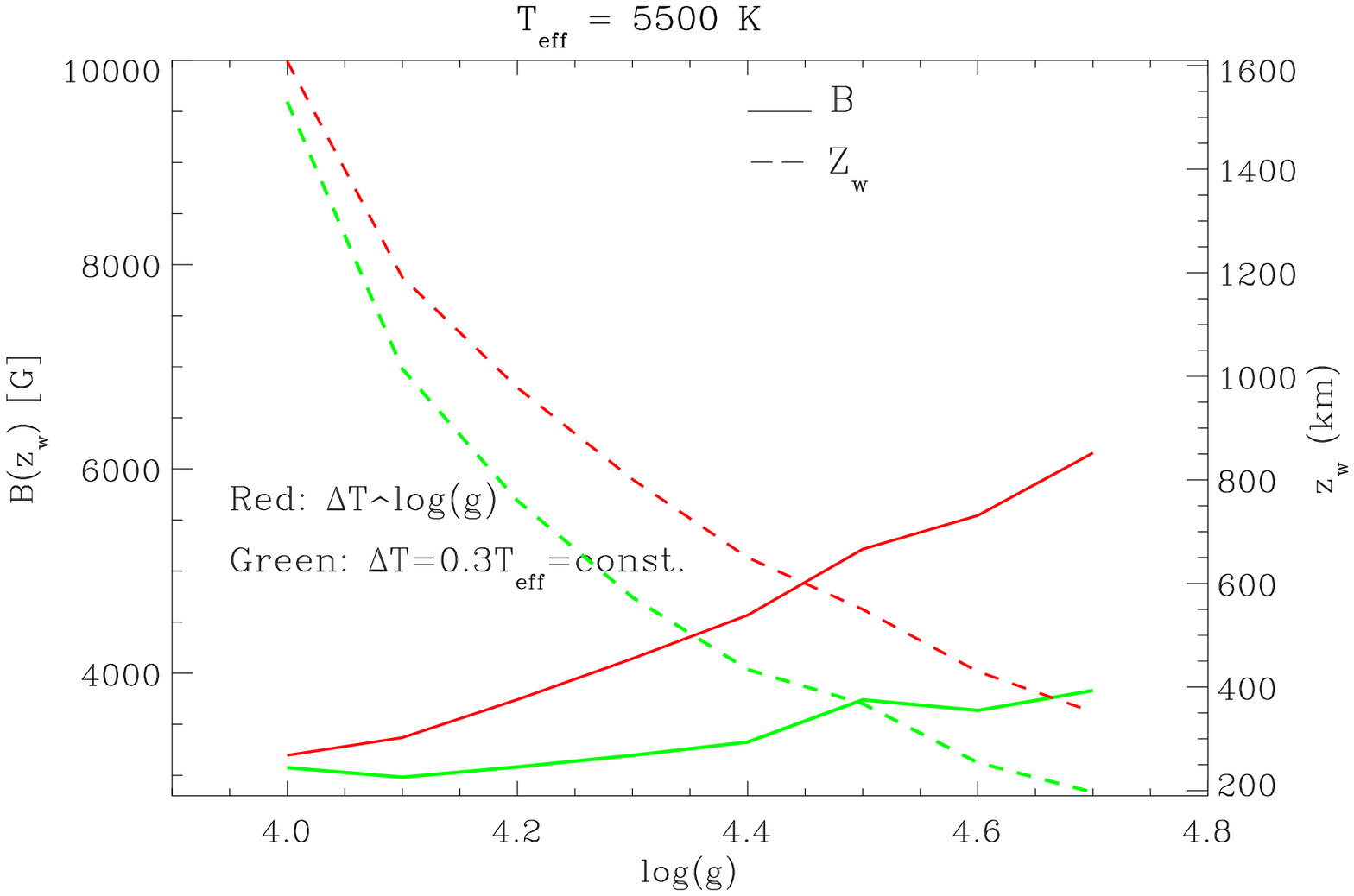}
 \includegraphics[width=3.5in]{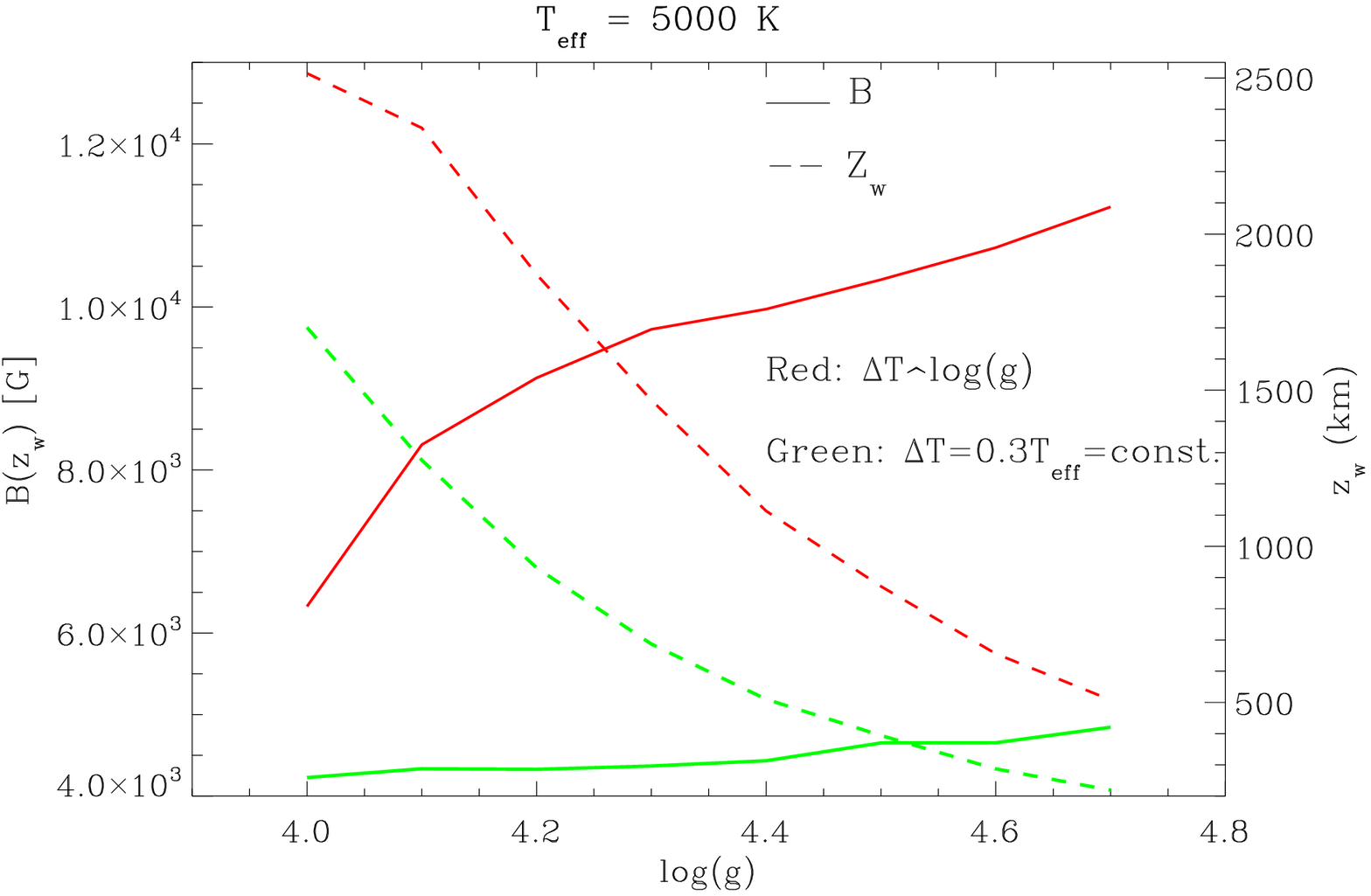}
 \caption{Same as Fig. 1 but for $T_{eff}=5500 K$ (top) and $5000 K$ (bottom)}
   \label{fig2}
\end{center}
\end{figure}
Figures 1 and 2 summarise the main results of our study. Each
figure shows the variation of observable field strengths $B(Z_{w})$ on the spot axis
as a function of $log (g)$ for a particular $T_{eff}$ for
the two distinct kinds of variation of spot temperature given as case (i) ({\it red curves}) and 
(ii) ({\it green curves}) in the previous Section.
The Wilson depressions, $Z_{w}$, are shown as dashed curves and the $B$ values
as solid curves. $Z_{w}$ are scaled in terms of pressure scale heights $H_{p}$
at the $\tau_{c}$=1 level in the quiet photospheres.
Fig.1, for $T_{eff} = 6000 K$, shows that the $B$ values are not very different
for the two cases above, and moreover the Wilson depressions $Z_{w}$ are of the
order of scale heights.
In contrast, Figure 2, for $T_{eff} = 5000 K$, shows that the results for the
above two cases are very different: the gravity dependent temperature contrasts
(as implied by the observed relation of O'Neal et al.(1996)) requires very
strong magnetic fields for starspot equilibrium, which stems from the large
values of $Z_{w}$. On the other hand, a constant
value of temperature reduction (case ii above) over the range of surface gravity
values used yields $B$ values within the observed ranges (see Berdyugina (2005) and references therein).
Consequently, a relatively less temperature reduction and hence a less amount of gas
evacuation is sufficient to attain a given value of field strength in cooler stars than
that in stars hotter than about $5500 K$. In other words, spots of a given field strength
would appear much darker in a star of $T_{eff}=6000 K$ than those in $T_{eff}=5000 K$.

According to Lockwood et al.(1992), younger and faster rotating stars
are 'spot dominated', i.e. more flux in spots than small-scale fields and faculae, and hence grow darker as activity increases. 
If such a 'spot domination' is purely dependent on age (rotation) but not on color (spectral type),
then it could be thought of as a phenomenon not contradictory with the $g$ dependent temperature contrasts
for spots derived by O'Neal et al. (1996). 
However, our results in Figure 2
imply unrealistically large $B$ and $Z_{w}$ values for spots in such cases.
On the other hand, consistency between Lockwood et al.'s results requiring
'spot domination' for younger stars and the situation of case (ii) of
our results requires that there be a color $T_{eff}$ dependence of spot properties
in addition to the age dependence.

Alternatively, results of Lockwood et al.(1992), but without the requirement of
'spot domination' in younger stars, could be made consistent with our case (ii) of less gas
evacuated spots if the small-scale fields forming the 'faculae' too are of such less evacuated
state. This would imply that faculae in younger and cooler stars are less bright and therefore
these stars grow darker as activity increases. Interestingly, the superadiabaticity that drives 
the convective collapse of small-scale flux tubes indeed linearly decrease with $T_{eff}$ and 
is found not to intensify such fields to a high degree of evacuation as in stars hotter than about
5000 K (\cite{rajaguruetal02}). Hence, it would appear that the near-surface thermal structure of
cool stars crucially determine the key properties of magnetic structures small and large. 
We conclude that the scaling of thermal and magnetic properties of starspots with
both the gravity $g$ (age) and $T_{eff}$ (color) are crucial for a consistent
interpretation of observed correlations between activity measures that sample the different
heights in the outer atmospheres.

\acknowledgements The presentation of this paper in the IAU Symposium 273 was possible due to  partial support  from the  National Science Foundation grant numbers ATM 0548260, AST 0968672 and NASA - Living With a Star grant number 09-LWSTRT09-0039.


\begin{thebibliography}{}
\bibitem[Alexander \& Ferguson (1994)]{alexander_ferguson94} {Alexander, D. R., \& Ferguson, J. W.} 1994, \textit{ApJ}, 437, 879
\bibitem[Berdyugina (2005)]{berdyugina05} {Berdyugina, S.} 2005, \textit{Living Rev. Solar Phys.}, 2, 8; \textit{URL: 
www.livingreviews.org/lrsp-2008-8}
\bibitem[Cox \& Giuli (1968)]{coxgiuli68} {Cox, J.P. \& Giuli, R. T.} 1968, \textit{Principles of Stellar Structure, Vol. 2}, 
New York: Gordon \& Breach, p590
\bibitem[Kurucz (1993)]{kurucz93} {Kurucz, R. L.} 1993, \textit{ATLAS9 Stellar Atmosphere Programs and 2 km/s grid, 
Kurucz CD-ROM No.13. Cambridge, Mass.: Smithsonian Astrophysical Observatory.}
\bibitem[Kurucz (2001)]{kurucz01} {Kurucz, R.L.} 2001, \textit{private communication}
\bibitem[Lockwood \etal\ (1992)]{lockwoodetal92} {Lockwood, G.W., Skiff, B.A., Baliunas, S.L., \& Radick, R.R.} 1992, 
\textit{Nature}, 360, 653
\bibitem[Maltby (1977)]{maltby77} {Maltby, P.} 1977, \textit{Solar Phys.}, 57, 335
\bibitem[O'Neal \etal\ (1996)]{onealetal96} {O'Neal, D., Saar, S. H., \& Neff, J. E.} 1996, \textit{ApJ}, 463, 766
\bibitem[Rajaguru \etal\ (2002)]{rajaguruetal02} {Rajaguru, S.P., Kurucz, R.L., \& Hasan, S.S.} 2002, \textit{ApJ}, 565, L101
\bibitem[Saar (1990)]{saar90} {Saar, S. H.} 1990, in \textit{IAU Symposium No. 138 Solar Photosphere: Structure, 
Convection, and Magnetic Fields}, ed. J. O. Stenflo (Kluwer: Dordrecht), p427
\bibitem[Solanki \etal\ (1993)]{solankietal93} {Solanki, S. K., Walther, U., \& Livingston, W.} 1993, \textit{A\&A}, 277, 639

\end{thebibliography}
\end{document}